# New designs of resistive microstrip gaseous detectors (R-MSGCs)

P. Martinengo[1], E. Nappi[2], R. Oliveira[1], V. Peskov[1,3], P. Pietropaolo[4], P. Picchi[5]

[1]CERN, Geneva, Switzerland
[2]INFN Bari, Bari, Italy
[3]UNAM, Mexico
[4]INFN Padova, Padova, Italy
[5]INFN Frascati, Frascati, Italy

**Abstract**

A new family of spark-protected micropattern gaseous detectors is introduced: a 2-D sensitive restive microstrip counter and hybrid detectors, which combine in one design a resistive GEM with a microstrip detector. These novel detectors have several important advantages over other conventional micropattern detectors and are unique for applications like the readout detectors for dual phase noble liquid TPCs and RICHs.

## I. Introduction

In the last few years, many efforts have been done by various groups to develop spark-protected micropattern gaseous detectors equipped with resistive electrodes instead of metallic ones [1]. A great success has recently been achieved with resistive GEM [2-6], resistive MICROMEGAS [6, 7] and resistive Well/CAT detectors [8-11].
In the recent reposts [1,12,13] we described spark-protected microstrip gas counters with resistive electrodes. We named them R-MSGCs. In this paper we present our new prototypes of such detectors: a 2-D sensitive R-MSGC and hybrid detectors combining in one design a GEM and a R-MSGC
. These innovative detectors are manufactured on standard PCB boards by using a simple technology, thus reducing the production cost. These novel detectors have several important advantages over other micropattern detectors and are unique for applications like the readout detectors for dual phase noble liquid TPCs and RICHs

## II. Detectors manufacturing
### II.1 2-D R-MSGC

This new detector type was manufactured from a conventional FR4 multilayer board 0.5 mm thick the top surface of which was coated with a 5 μm thick Cu layer (see Fig.1a). Two bottom layers of this board consisted from 0.1mm thick FR-4 sheets each equipped with parallel metallic readout strips; the width of the readout strips was 200 μm and their pitch was 1mm. The strips of the second layer (from the top) were oriented perpendicular to the strips of the third layer. Subsequently, on the top surface of the PCB, parallel grooves were milled 100 μm deep, 0.6 mm wide, with a pitch of 1mm. These grooves were oriented parallel to the strips of the third layer (see Fig.1b). The grooves were then filled with a resistive paste (ELECTRA Polymers) and the R-MSGC surface was chemically cleaned (see Fig. 1c). Finally, by using a photolithographic technology, Cu 20 μm wide strips were created between the grooves (Fig 1d). Then this detector was glued on a 2mm thick FR-5 supporting plate (Fig. 1e). This technology is simpler than that used before for manufacturing prototypes #1 and #2 [12, 13] and consequently the cost of the detector is cheaper. Both the anode and the cathode strips were covered near their edges by a Coverlay layer to avoid surface discharges (see Fig.2). The total resistivity of each cathode strips was 300 MΩ and their resistivity in the region C (see Fig. 2) was 100MΩ.

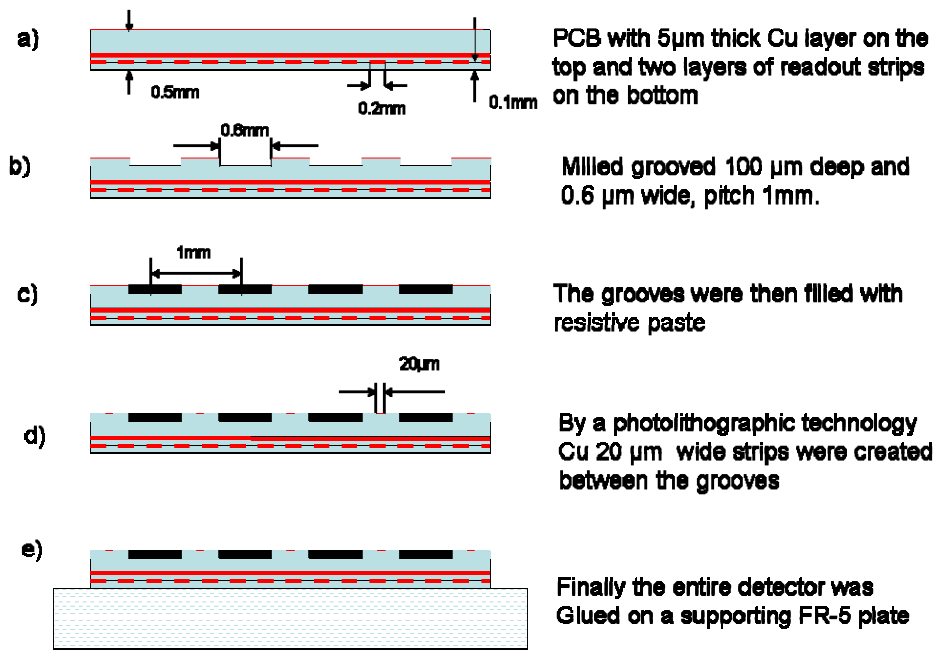

Fig.1. Schematic illustration of a R-MSGC manufacturing process

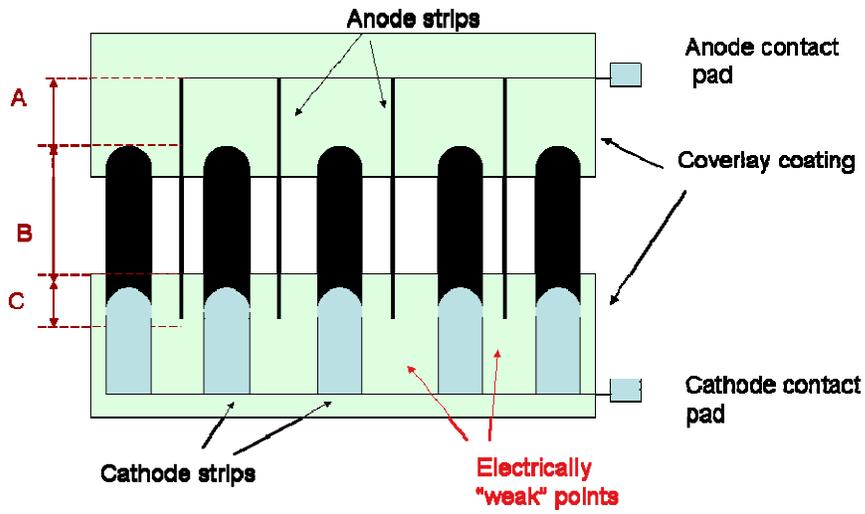

Fig. 2. A schematic drawing of the R-MSGC

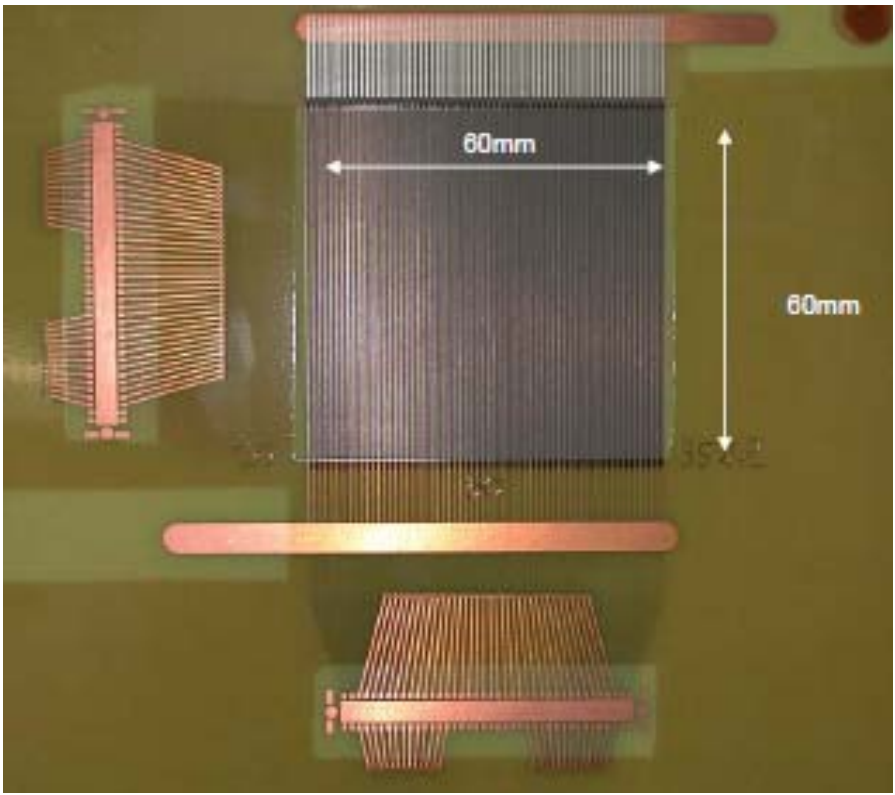

Fig.3. A photograph of a 2-D sensitive R-MSGC having capacitive coupled readout strips

The photograph of the R-MSGC is shown in Fig. 3 . In the centre of this photograph one can see the active area of the R-MSGC: its anode and cathode strips. On the right side of the photo and on its bottom are shown the rows of readout strips located under the R-MSGC

**II.2 Hybrid detectors**

Two types of hybrid detectors were manufactured. One was an R-MSGC combined with a GEM in one design. The concept of this detector is resembling the so called MHCP detector [14], however the important differences were that it was manufactured from a printed circuit plate 0.4 mm thick and had resistive cathode strips (see Figs. 4-6) making it spark-protective. The manufacturing procedure was similar to that used in the production of the R-MSGC (see for example [13]), with an additional last step which includes holes drilled by a CNC machine. The geometrical characteristics of this detector were: anode strips width 20μm, cathodes width 0.6mm, pitch 1mm, a hole diameter 0.3mm, active are 60x60mm$^2$.
The second detector was quiet similar to the Thick -COBRA design [15], but featuring a resistive back plane instead of a metallic one (Fig.7).

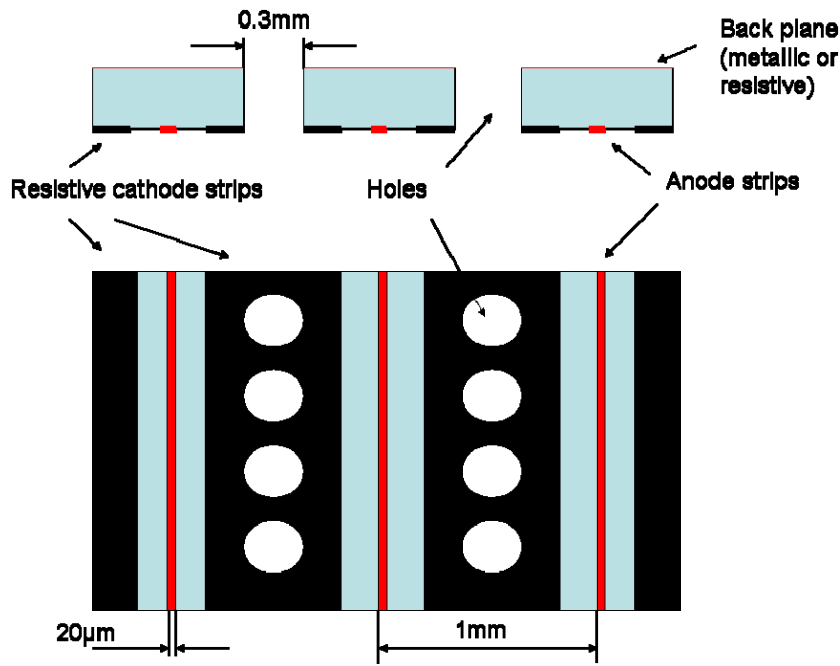

Fig. 4. A schematic drawing of the resistive hybrid detector

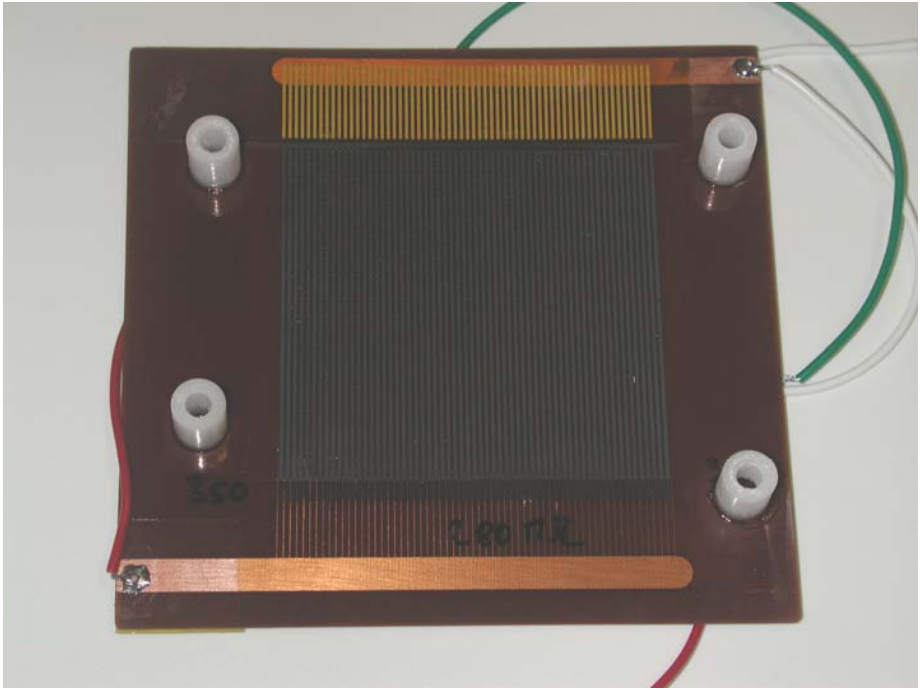

Fig.5. A photograph of a hybrid R-MSGC

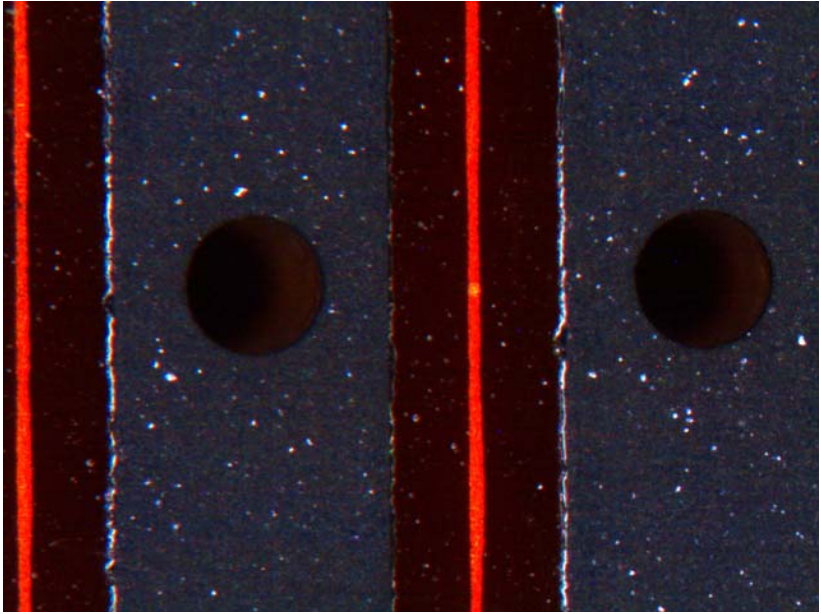

Fig.6.A magnified photo of the active region of the hybrid R-MSGC

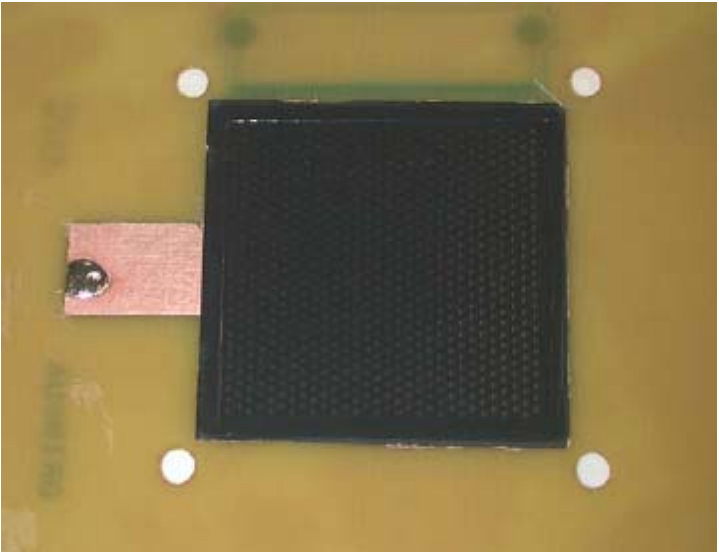

Fig.7. A photograph of Thick COBRA with resistive electrodes

### III. Experimental setup

The detector characteristics were measured with the setup used in [16]-see Fig 8. The gas chamber was installed inside a cryostat allowing, if necessary, the measurements at temperatures below the room one. In some tests, the $CaF_2$ window was replaced by a beryllium window combined with a lead collimator. Tests were performed in Ar and mixtures of Ar with $CH_4$ and $CO_2$ in the range of a quencher concentration between 10 and 20%.

The primary ionization in the detector volume was produced by $^{241}$Am (alpha particles) and by $^{55}$Fe source (6 keV photons). For position resolution measurements we used the $^{55}$Fe source combined with a slit collimator, 0.3mm with.

In some applications, it has also been tested the operation of R-MSGC combined with a CsI coated TGEM operating as a preamplifaction structure. The latter had thickness of 0.45 mm, hole diameter 0.4 mm, rims thickness 10 µm and 0.8 mm pitch.

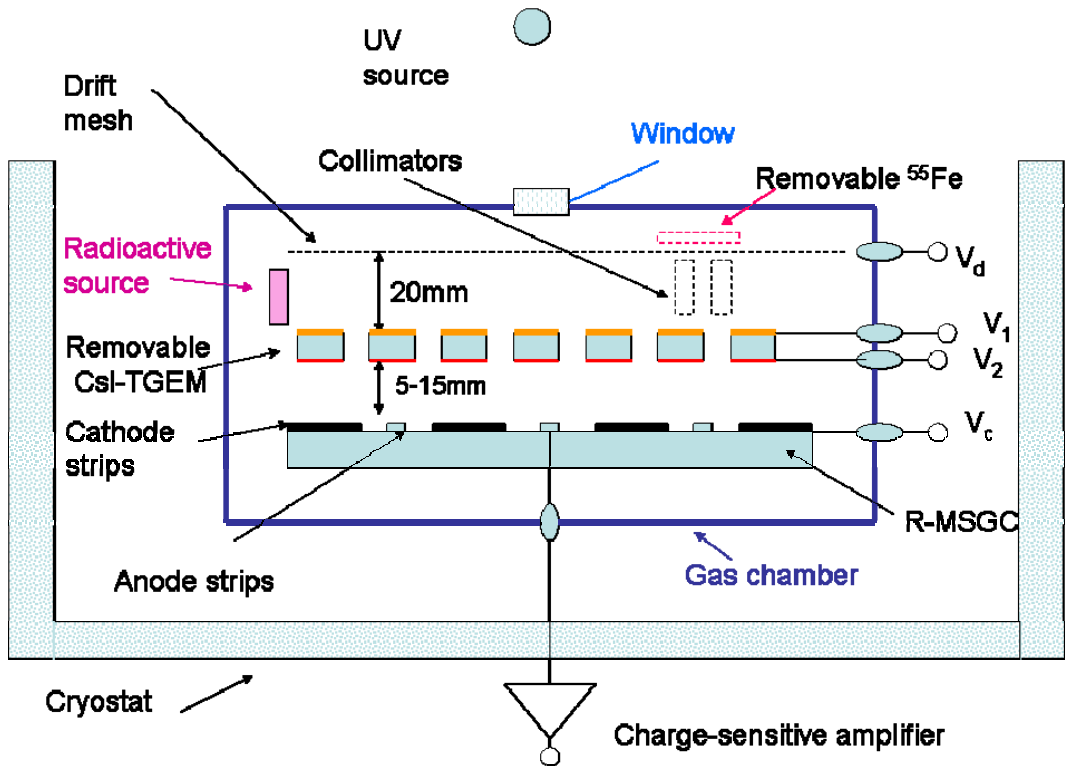

Fig.8. A schematic drawing of the experimental set up used for tests of R-MSGC and hybrid detectors having resistive electrodes

## IV. Results
### IV.1 2-D R-MSGC

Figures 9,10 show curves of the R-MSGC gain as a function of the voltage measured in Ar and in some mixtures of Ar with $CH_4$ and $CO_2$. As can be seen in all these gases *the* maximum achievable gain was $10^4$- as high as it can be achieved with the best MSGC manufactured on glass substrate. However, in contrast to classical MSGCs, R-MSGCs, due to the resistivity of electrodes and the small capacity between the strips, the spark energy was strongly reduced so the strips were not damaged even after many sparks. The energy resolution measured in Ar+$CH_4$ and Ar+$CO_2$ gas mixtures was about 25% FWHM for 6 keV photons.
Results of measurements of induced signals from the readout strips are shown in Fig.9. Although the profile of the induced signals in our particular conditions is comparable to the readout strips pitch size, changes in the position collimator on 200μm are noticeable. The position relation estimated from the centre of gravity is about 200 μm. More precisely the position resolution will be determined during the beam test performed with charged particles.

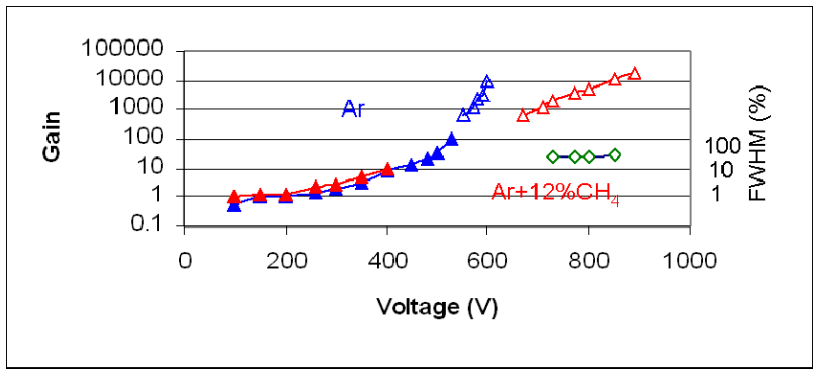

Fig.9. Gain (triangles) and energy resolution (rhombuses) dependence on voltage applied to R-MSGC measured in Ar (blue symbols) and Ar+12%$CH_4$ (red symbols). Filled triangles-measurements performed with alpha particles, open triangles - $^{55}$Fe.

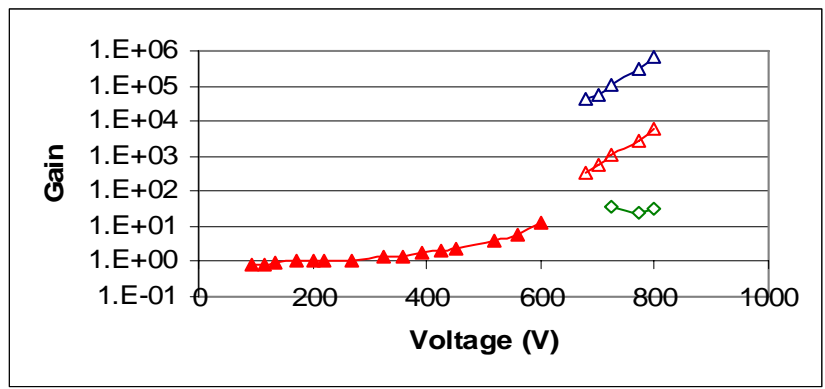

Fig.10. Gas gain curves measured in Ne+10%$CO_2$: filled triangles –alpha particles, open symbols- $^{55}$Fe. Open rhombuses-energy resolution. Blue triangles represent gas gain measures with a CsI-coated TGEM preamplification structure.

## IV.2 Hybrid R-MSGC

Original MHCP and COBRA detectors were developed with the aim to reduce the avalanche ion back flow and are oriented on applications as TPC or gaseous photodetectors [15]. Obviously, our hybrid detectors with resistive electrodes can also be used in these applications, however the main focus of our work is on cryogenic TPCs and dark matter detector (see next paragraph). For this reason we have mainly studied the operation of the hybrid detector at low temperatures. As an example, in Fig.12 are shown gas gain curves measured with the hybrid R-MSHC at 1atm at room and low temperatures

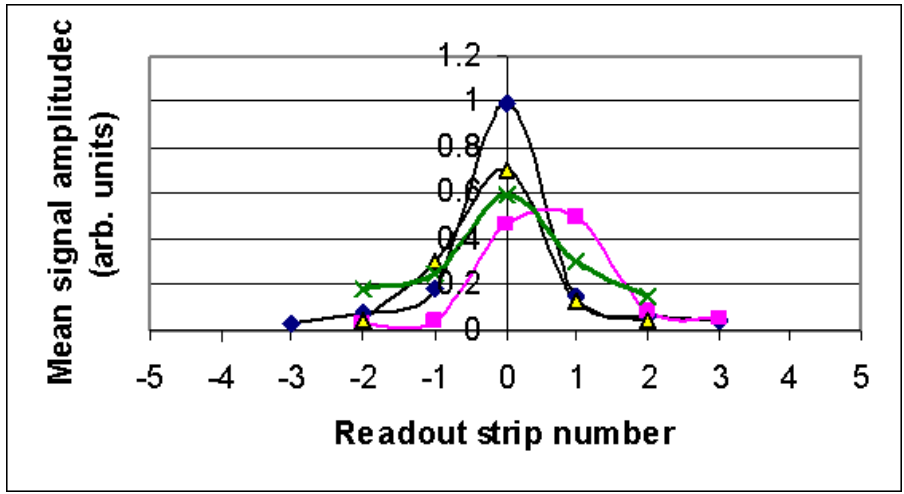

Fig.11. Results of measurements induce signals from the readout strip oriented along (green curve with crosses) and perpendicular to the anode strips of R-MSGCs (rhombuses, triangles and squares). Rhombus- the collimator is aligned along the strip #0. Triangles -the collimator was moved on 200μm towards the strip#1. Squares- the collimator was aligned between the strip#0 and # 1. Measurements were performed in Ar+10%$CO_2$ at a gas gain of $5 \times 10^3$.

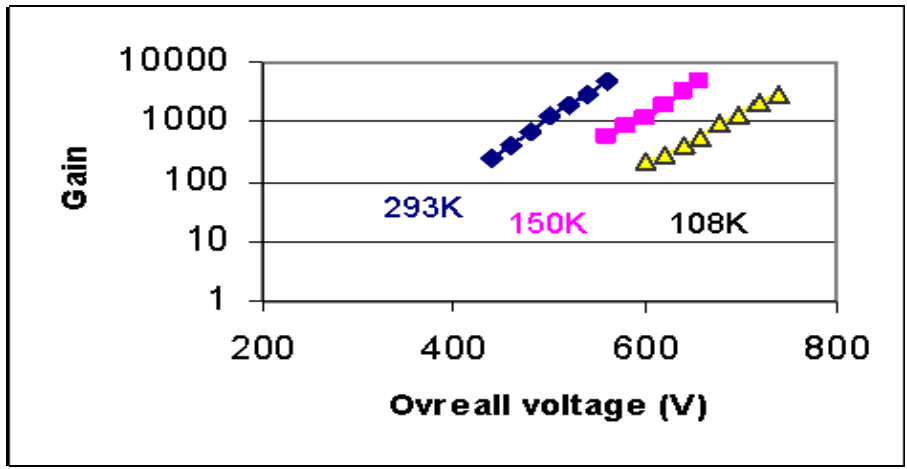

Fig.12. Gas gain vs. the overall voltage (the voltage applied across the holes and between the anode and cathode strips) for hybrid R-MSGC measures in Ar at various temperatures

As can be seen gains above $10^3$ were achieved even at 108K, when the gas density was almost three times higher compared to room temperature

## V. Possible applications:

The gas gains achieved with our 2D sensitive R-MSGCs (~$10^4$) are comparable to those obtained with spark-protected MICROMEGAS having resistive anode strips [7]. Currently our group is studying the feasibility of building a large- area RICH detector (~1x1$m^2$) consisting from a CsI-coated TGEM preamplification structure and a 2-D sensitive R-MSGC. Such a photodetector can be an interesting option for a so-called VHMPID detector which is under the study in the frame of the ALICE upgrade program [17]. A 5x5$cm^2$ prototype was already assembled and tested (see Figs.8 and 10) and first promising results we obtained: we were able to detect simultaneously beta particles produced by $^{90}$Sr (simulating minimum ionizing particles background in real experiment) and single UV photons produce by a UV lamp (simulating Cherenkov photons).A larger-20x20$cm^2$ - prototype is already under the construction now.

As was described in [12] the spark-protected COBRA will be an attractive option for the detection of charge and light from the LXe TPC with a CsI photocathode immersed inside the liquid.

## VI . References: